\documentclass[aps, prl, twocolumn, 10pt, nofootinbib]{revtex4-1}
\usepackage[sort&compress]{natbib}
\usepackage{amsmath,amssymb,amsfonts}
\usepackage{graphicx}
\usepackage{subfigure}
\usepackage{xcolor}
\usepackage[pagebackref,colorlinks=true]{hyperref}
\usepackage[UKenglish]{babel}
\usepackage[UKenglish]{isodate}
\usepackage{multirow}
\usepackage{braket}
\usepackage{enumitem}
\usepackage{pdfpages}
\usepackage{etoolbox}
\usepackage{lipsum} 
\usepackage{float}

\newcommand{\nit}[1]{\noindent\textit{#1}}

\makeatletter
\patchcmd{\@outputpage@head}{\@ifx{\LS@rot\@undefined}{}{\LS@rot}}{}{}{}
\makeatother
\definecolor{citeblue}{HTML}{0E5484}
\hypersetup{colorlinks=true,citecolor=citeblue}

\begin{document}
\title{A Single-Photon-compatible Telecom-C-Band Quantum Memory in a Hot Atomic Gas}
\author{S. E. Thomas$^{1}$, S. Sagona-Stophel$^{1}$, Z. Schofield$^{2}$, I. A. Walmsley$^{1}$, P. M. Ledingham$^{2}$}
\email{P.Ledingham@soton.ac.uk}
\affiliation{$^1$QOLS, Department of Physics, Imperial College London, London SW7 2BW, UK\\
$^2$Department of Physics and Astronomy, University of Southampton, Southampton SO17 1BJ, UK}
\date{\today}
	
\begin{abstract}
The efficient storage and on-demand retrieval of quantum optical states that are compatible with the telecommunications C-band is a requirement for future terrestrial-based quantum optical networking. Spectrum in the C-band minimises optical fiber-propagation losses, and broad optical bandwidth facilitates high-speed networking protocols. Here we report on a telecommunication wavelength and bandwidth compatible quantum memory. Using the Off-Resonant Cascaded Absorption protocol in hot $^{87}$Rb vapour, we demonstrate a total memory efficiency of $20.90(1)\,\%$ with a Doppler-limited storage time of $1.10(2)\,$ns. We characterise the memory performance with weak coherent states, demonstrating signal-to-noise ratios greater than unity for mean photon number inputs above $4.5(6)\times10^{-6}$ per pulse.
\end{abstract}
	
\maketitle

\nit{Introduction} -- Quantum light-matter interfaces are a critical component for quantum repeating \cite{Briegel1998,Gisin2007} technology and therefore building large-scale quantum networks \cite{Ladd2010, Komar2014,Wehner2018}. Such interfaces allow for a quantum memory \cite{Heshami2016}, a device capable of storing true single photons with both high fidelity and efficiency, with on-demand retrieval after a sufficiently long storage time. Besides these important attributes, compatibility with wavelengths in the telecommunications C-band, as well as GHz operational bandwidth, is of crucial importance to allow high repetition rate quantum photonic operations in low-loss optical fiber networks.

Significant progress has been made realising telecom quantum optical memories across different material platforms. Atomic-based systems require an optical transition in the telecom range, with cryogenically-cooled trivalent erbium ions in solid-state hosts a popular choice \cite{Sun2005}. Protocols including the Atomic Frequency Comb (AFC) \cite{Lauritzen2011, Craiciu2019, Stuart2021}, stark-modulated AFC \cite{Craiciu21, Liu2022}, controlled-reversible inhomogeneous broadening \cite{Lauritzen2010} and revival of silenced echo \cite{Dajczgewand2014} have been performed in erbium doped Y$_2$SiO$_5$, which include realisations in nano-photonic structures \cite{Craiciu2019, Craiciu21}. Other hosts have been explored, with demonstrations of AFC in erbium-doped optical fiber \cite{Jin2015, Wei2022} and lithium niobate waveguides \cite{Askarani2019}. Further, frequency conversion via non-linear processes have been used to interface telecom light with quantum memories, including four-wave mixing in cold atomic clouds \cite{Radnaev2010, Ding2012} and non-linear crystals \cite{Maring2014, Luo2022}. Furthermore, opto-mechanical interactions have enabled telecom light storage in optical fiber \cite{Zhu2007}, chalcogenide glass chips \cite{Merklein2017} and in nanofabricated mechanical resonators \cite{Wallucks2020}. 

Despite these impressive advances, the combination of on-demand read-out with high signal-to-noise ratio (SNR) remains elusive. Erbium-based systems suffer poor optical pumping leading to high absorbing backgrounds that limit the overall memory efficiency \cite{Lauritzen2010}, requiring magnetic fields of the order of $7\,$T and a $2\,$K temperature to overcome \cite{Stuart2021}. Frequency conversion adds layers of inefficiency and in the case of non-linear crystals, adds noise via pump-induced Raman scattering and spurious spontaneous parametric down-conversion \cite{Pelc2011}. Opto-mechanical systems need to overcome the challenge of reaching the mechanical quantum ground state to remove thermal-phononic-mode induced noise \cite{Shandilya2021}. In this work, we circumvent these problems and demonstrate an ultra-low-noise, high bandwidth quantum memory using a telecom transition in hot rubidium vapour.

\begin{figure}[hb]
\centering\includegraphics[width=\linewidth]{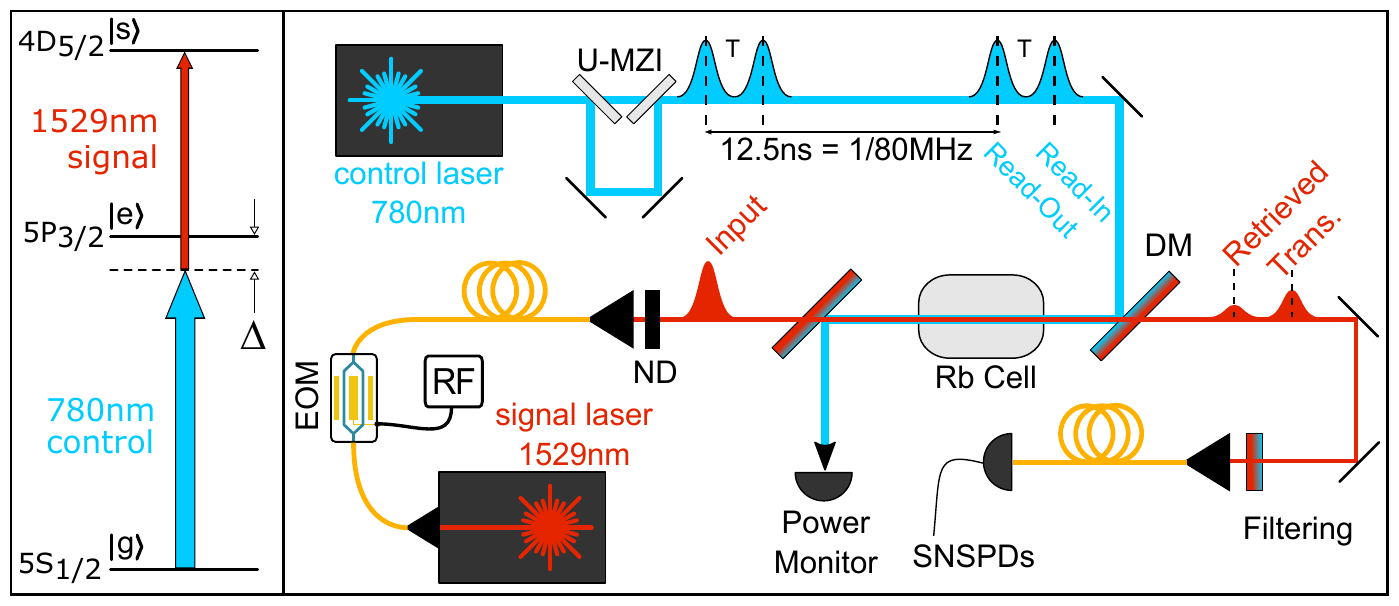}
\caption{Left Panel: Telecom ORCA level scheme. Right Panel: Experimental Set-up.  DM: Dichroic Mirror; ND: Neutral density filters; RF: Arbitrary Waveform Generator; SNSPD: Superconducting Nanowire Single Photon Detectors; U-MZI: Unbalanced Mach–Zehnder interferometer.\label{ORCA}}
\end{figure}

\nit{Protocol} -- We use the off-resonant cascaded absorption (ORCA) \cite{Kaczmarek2018} protocol for our demonstration. ORCA comprises two counter-propagating optical fields that are two-photon resonant with an optical ladder transition of an atomic ensemble while being detuned from the intermediate transition by $\Delta$ (Fig. \ref{ORCA}), giving rise to absorption of the beams by exciting the atoms in the ensemble. The intense control pulse dresses the atomic system providing a broad virtual state to which the signal field couples efficiently, so that it may be completely absorbed. For large enough intermediate detuning, the absorption bandwidth is determined by the bandwidth of the control pulse. The input pulse excites a superposition of atoms in the ensemble of the $|\textrm{g}\rangle - |\textrm{s}\rangle$ transition. The state of the atomic ensemble after absorption of the signal is of the form $\frac{1}{\sqrt{N}} \sum_{j=1}^N e^{i(\vec{k}_s - \vec{k}_c)\cdot\vec{v}_jt}\ket{g_1,\dots,s_j,\dots,g_N}$ where $\ket{g(s)_j}$ labels the ground (storage) state of the $j^\mathrm{th}$ atom,  $N$ is the total number of atoms, $\vec{k}_{s(c)}$ is the input (control) pulse wave vector, $\vec{v}_j$ is the velocity of the $j^\mathrm{th}$ atom. The collective dipole moment will emit light upon the application of a second control pulse  {after a storage time $T$}, allowing the signal to be read out and completing the memory operation.

The main benefits of ORCA are threefold. Firstly, the optical fields can be at significantly different wavelengths, depending on the chosen atomic species and on the levels used. This makes it simple to filter the strong control from the single-photon-level signal. Secondly, the storage state is a doubly-excited one and so has no atomic population under normal operational conditions. This simplifies the approach by not needing any initialisation of the atomic ensemble with optical pumping fields, and eliminates spontaneous emission noise associated with inefficient optical pumping. Finally, four-wave mixing type noise is eliminated. For example, in Lambda-based approaches, the strong control is able to efficiently Raman scatter from the populated ground state, spontaneously emitting a photon and leaving spurious excitations in the storage state that are subsequently read-out into the same spatial-spectral-temporal mode as that of the desired signal \cite{Thomas2019}. Equivalent processes with a ladder system are not possible. In the case of a non-degenerate ladder system, there is no two-control-photon process that can populate the storage state and so the ORCA protocol is inherently noise-free in this respect.

ORCA has been achieved in hot atomic vapours demonstrating low-noise operation over GHz bandwidth allowing for high SNR for single-photon-level inputs \cite{Finkelstein2018}, recall of heralded single photons \cite{Kaczmarek2018}, and can be used as a time-nonstationary coherent spectral-temporal filter to improve indistinguishably of single photon sources \cite{Gao2019}. {Here we extend the ORCA protocol to store light in the telecom C-band for the first time.}

\nit{Experimental Set-up} -- See Fig. \ref{ORCA}. An ensemble of rubidium atoms in a cell of length $8\,$cm  is heated to around $120^\circ$C yielding more than $10^9$ atoms into the vapour phase. With spectroscopy we estimate the vapour to contain $96.9\,\%$ of the isotope $^{87}$Rb. The ladder system used for this demonstration is depicted in the left panel of Fig.~\ref{ORCA}. We utilise the doubly-excited $4$D$_{5/2}$ state as the storage state $|s\rangle$, that connects to the ground state $|g\rangle$ ($5$P$_{1/2}$, F$=2$) via a two-photon process with the telecommunication C-band compatible wavelength $1529.3\,$nm (signal) and the Rb D2 line at $780.3\,$nm (control). The signal (control) field is blue- (red-) detuned from the intermediate level $|e\rangle$ ($5$P$_{3/2}$, F$'=\{1,2,3\}$) by $\Delta = 6\,$GHz.

Control pulses of a few nJ energy are delivered by a mode-locked Titanium Sapphire laser (Spectra Physics Tsunami) at a repetition rate of $80\,$MHz. The pulse bandwidth is approximately $1\,$GHz. An unbalanced Mach–Zehnder interferometer (U-MZI) splits an individual pulse into the read-in and read-out pulses, separated by time $T$. We use polarising beamsplitters to construct the U-MZI, with a half waveplate before used to adjust the energy into each of the pulses. The output of the U-MZI is polarized by a half waveplate and PBS (not shown),  {to ensure that the read-in and read-out control pulses have the same polarization}. Dichroic mirrors are used to overlap and separate the control and the signal mode, and we monitor the average power of the control pulses with a power meter (Thorlabs). The signal is generated from a CW telecom laser (Santec) that is passed through an intensity modulator (iXBlue) driven by a fast arbitrary waveform generator (Tektronix) to generate Gaussian pulses of around $300\,$ps duration synchronised to the repetition rate of the control pulse laser and delivered at a rate of $10\,$MHz. Neutral density filters are used to reduce the intensity of the pulses to the single-photon level. High transmission band- and long-pass filters are placed before the detection fiber to provide 14 orders of magnitude suppression of background light at the control pulse wavelength. Photons are detected with superconducting nanowire single photon detectors (Photon Spot) together with a time-to-digital converter (Swabian) to produce start-stop histograms.

\begin{figure*}[htpb]
\subfigure[]{\includegraphics[width=0.6\textwidth]{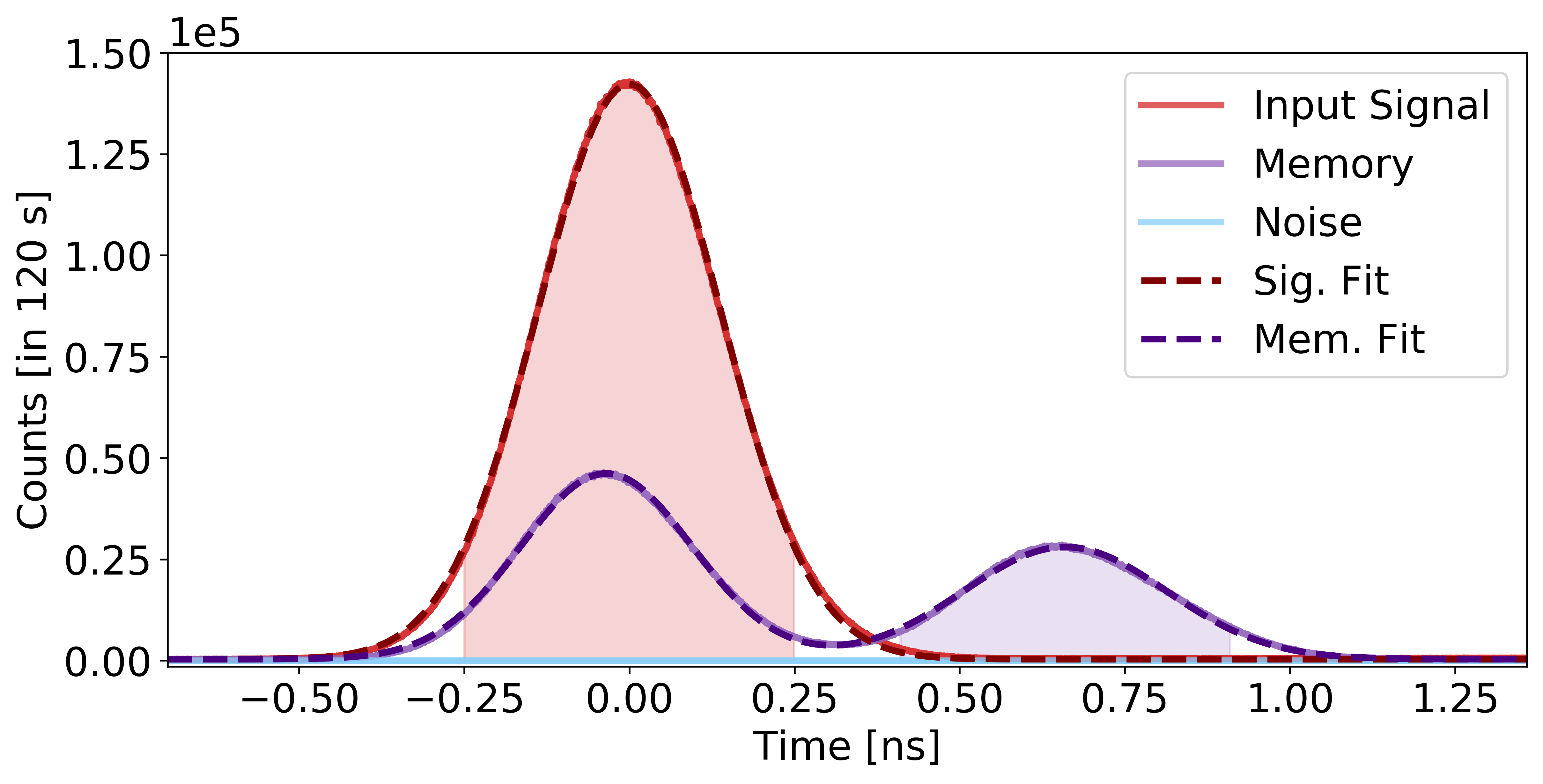}\label{CountsVsTime}}
\subfigure[]{\includegraphics[width=0.3\textwidth]{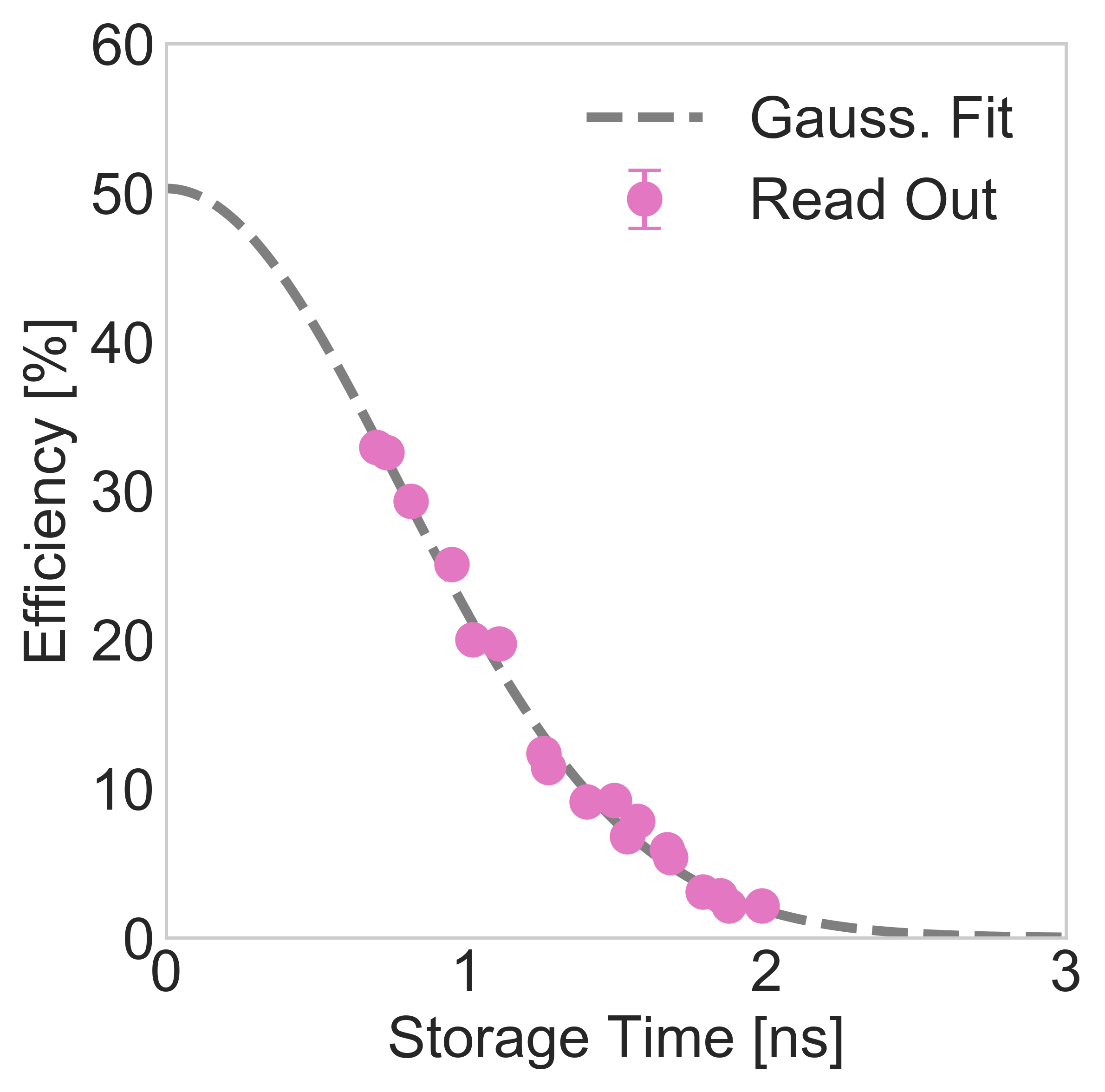}\label{Tau}}
\caption{(a) Storage and retrieval of telecom light. The solid lines are the start-stop histograms for the signal at input (red), at the memory output (purple) and the noise without signal (light blue) - the timing resolution of the TDC is $1\,$ps and the measurement time for each data set is $120\,$s. The dashed lines show Gaussian fits to the measured signal. The shaded areas indicate the read-in and read-out integration windows of $500\,$ps width. The storage time in this case is $660\,$ps. (b) Read-out efficiency vs storage time. Pink circles are the data which are fit with a Gaussian function of control delay (grey dashed line). For this data, the measurement time per data point is $10\,$s and integration window as before.\label{ORCA_Data}}
\end{figure*}

\nit{Memory Performance} -- Fig. \ref{CountsVsTime} shows a plot of storage and recall of telecom light for the highest measured storage efficiency. We implement Gaussian fits to these traces and use these fits to calculate photon number and efficiencies. We use an input pulse with an average number of photons per pulse of $\mu_\textrm{in} = 0.084(10)$. The mean photon number was estimated by integrating the counts over a $500\,$ps integration window centred on the $350\,$ps FWHM pulse at $t=0$, then dividing out detection loss $(\eta_\textrm{det} = 80(8)\,\%)$ and transmission from the input of the cell to the detector $(\eta_\textrm{trans} = 56(4)\,\%)$, and finally normalising by the number of trials in the $120$s total measurement time. Two control pulses, with a time separation of $660\,$ps, are applied to read in and read out the signal to the memory, with respective energies of $0.57(4)\,$nJ and $3.6(3)\,$nJ. When photons are successfully stored, they are measured in the time-region called the ``read-out window" -- while photons that are not successfully stored pass through undelayed by the memory in the ``read-in  window". We define the read-in efficiency as the ratio of counts remaining in the read-in window when the control is applied to when it is not -- in this case it is $\eta_\textrm{read-in}\,=\,69.13(1)\,\%$. We see clear retrieved signal in the read-out window, defined as the $500\,$ps window centred at $t = 660\,$ps. The total memory efficiency, defined as the ratio of read-out to input counts, is measured to be $\eta_\textrm{mem} = 20.90(1)\,\%$, resulting in read-out efficiency of $\eta_\textrm{read-out}\,=\,30.23(1)\,\%$. The total throughput efficiency for the input at the front of the cell to being stored, retrieved and detected is $\eta_\text{mem}\eta_\text{trans}\eta_\text{det} = 9.4(1.2)\%$. 

\nit{Storage Dynamics} -- Fig. \ref{Tau} shows the read-out memory efficiency as a function of read-out delay after the storage. Performed under similar conditions to Fig. \ref{CountsVsTime}, we measure an average read-in efficiency of $\eta_\textrm{read-in} = 78(2)\,\%$ and infer the read-out efficiency at zero storage time from the fit to be $\eta_\textrm{read-out} = 50(2)\,\%$. The read-in and read-out efficiencies are not matched. Forward retrieval  {(i.e the retrieved signal propagates in the same direction as the input)} results in re-absorption of the signal toward the exit of the cell, reducing the maximum efficiency for fixed control pulse energy. Backward retrieval  can, in principle, avoid this \cite{Surmacz2008}. 

The characteristic lifetime of $1.10(2)\,$ns is extracted from a Gaussian fit. The lifetime is limited by inhomogenous Doppler broadening of the atoms causing dephasing of the atomic polarization. Atom $j$ constituting the collective dipole moment between the ground and storage state has a net wavevector $e^{i(\vec{k}_s - \vec{k}_c)\cdot\vec{v}_jt}$ where $(\vec{k}_s - \vec{k}_c)\cdot\vec{v}_j$ is the Doppler shift for the $j^\textrm{th}$ atom. Given the broad Maxwell-Boltzmann distribution of velocities of the atomic vapour, each atom accumulates a different phase thereby reducing the collective dipole moment. The spatial wavelength of the collective orbital-wave excitation is around $1.6\,\mu$m, a distance traversed by an atom in the hot vapour in a few nanoseconds.  The storage time is also reduced by the interference of multiple pathways through different hyperfine states with characteristic timescales of $\sim 10$ns \cite{Kaczmarek2018}. The natural lifetime of the storage state provides the ultimate limit of $\sim\,90\,$ns \cite{Theodosiou1984, Safronova2004}. The timescale we have measured here is consistent with that expected from Doppler dephasing \cite{Main2021} suggesting strongly that this is the main limitation for our storage time.

\begin{figure}[htbp]
\centering
\includegraphics[width=0.9\linewidth]{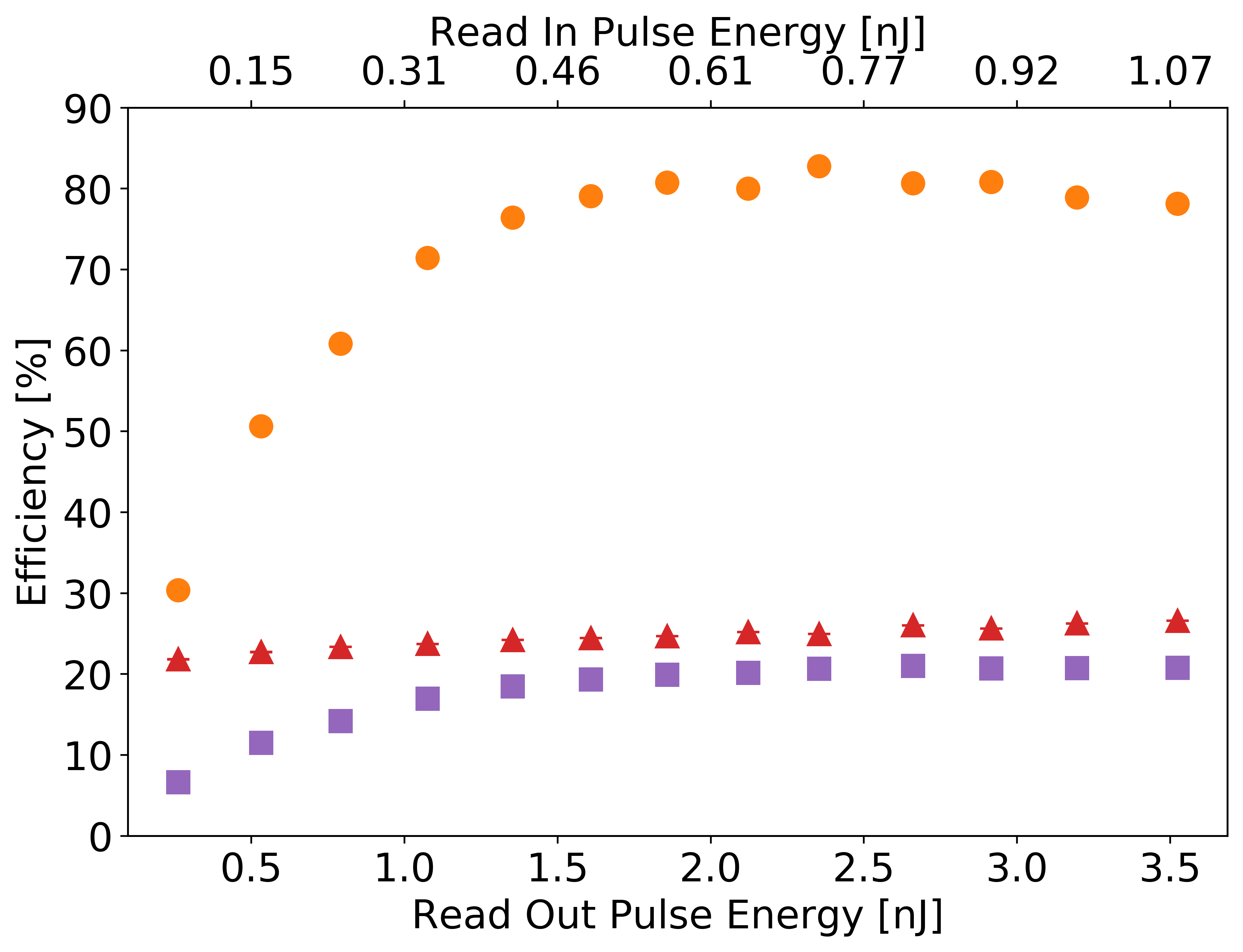}
\caption{ORCA memory vs pulse energy. Plot shows read-in efficiency (orange circles), read-out efficiency (red triangles) and total memory efficiency (purple squares) for different energies in read-in and read-out control pulses for $\mathcal{R} = 3.3(1)$. For this data, the measurement time per data point is $10\,$s and integration window $500\,$ps.\label{Energy}}
\end{figure}

\nit{Control Dependence} -- The data presented in Fig.~\ref{ORCA_Data} show the optimised output efficiency. The ORCA memory efficiency varies with the total energy in the control pulses, and is shown for a read-out to read-in ratio, $\mathcal{R}\,=\,E_{\textrm{R}_\textrm{out}}/E_{\textrm{R}_\textrm{in}}$ of $\mathcal{R}\,=\,3.3(1)$ in Fig.~\ref{Energy}. Note that this is a different condition for Fig.~\ref{ORCA_Data} with $\mathcal{R}\,=\,6.4(1)$, allowing for more pulse energy for the read-in process (and so less for the read-out). The read-in efficiency quickly increases and then begins to decrease when the read-in pulse energy is near $0.7\,$nJ. This decrease is the consequence of the AC-Stark effect by which the atomic resonances are frequency-shifted by the control pulse. The effect scales as $\Omega^2/\Delta$ where $\Omega$ is the Rabi frequency of the control field, and so higher control pulse energies result in shifts large enough to move the optical fields out of two-photon resonance and reduce the efficiency. One approach to prevent this is to optimise the spectral-temporal mode of the control pulse for a given pulse energy \cite{Nunn2007, Gorshkov2007}. This effect is not seen on the read-out efficiency because the optical mode that is read out will frequency shift with the atoms.

\begin{figure}[htbp]
\centering
\includegraphics[width=0.9\linewidth]{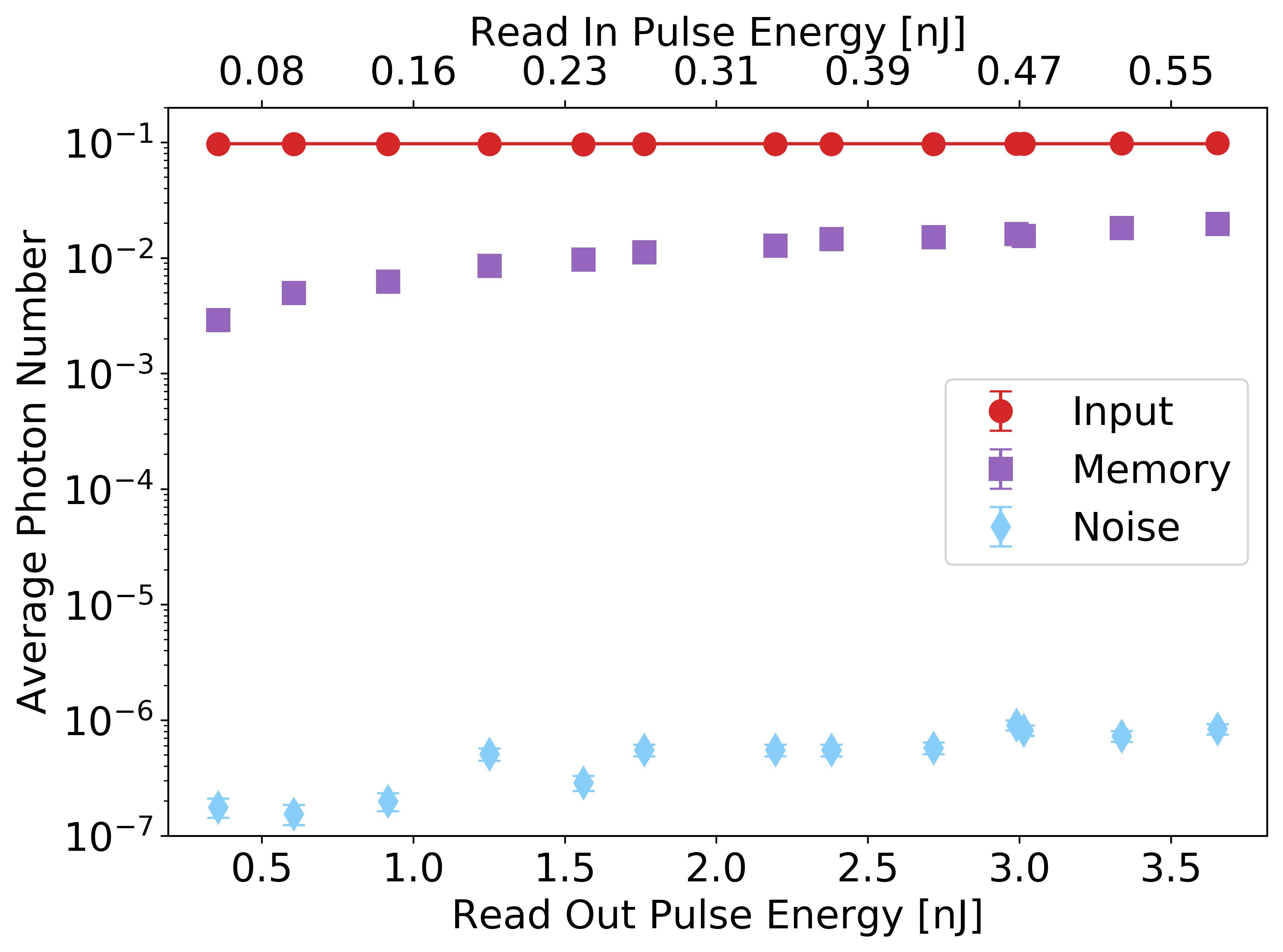}
\caption{The average photon number per integration bin for the input (red cricles), memory (purple squares) and noise (light blue diamonds) as a function of control pulse energy. For this data, $\mathcal{R}\,=\,6.4(1)$, the measurement time per data point is $10\,$s and integration window $500\,$ps.\label{AvPhotonNumber}}
\end{figure}

\nit{Noise Performance} -- A critical requirement for any quantum memory is that the output SNR be sufficiently high to retain the input quantum state upon readout. In Fig.~\ref{AvPhotonNumber} we assess the SNR by quantifying the average number of photons per pulse in the input, the retrieved output, and the noise floor (when the input signal is blocked but the control pulses are switched on) for $\mathcal{R}\,=\,6.4$. We measure the noise floor to be less than $10^{-6}$ photons per cycle.  There is a slight linear increase of the noise as a function of control pulse energy which we attribute to either leakage of the control pulse or leakage of one-photon scattering events of the control with the atoms. In spite of this we observe an SNR of around 4 orders of magnitude for every control pulse energy tested using $\mu_\textrm{in}$ at about $0.1$. For the data in Fig.~\ref{CountsVsTime} with input photon number $0.084$ we measure the SNR to be $1.9(1)\,\times\,10^{4}$. 

A useful metric to compare the performance of quantum memories is the ratio of the noise counts on the output to the total memory efficiency: $\mu_1 = N/\eta_\textrm{mem}$ \cite{Gundogan2015}. This parameter $\mu_1$ is essentially the number of photons per pulse on average at the input of the quantum memory that gives an SNR = 1 at the output. For the data in Fig. \ref{CountsVsTime} we measure the noise to be $N = 9(1) \times 10^{-7}$ photons per pulse over the integration window of $500\,$ps. This leads to $\mu_1$ = $4.5(6)\times10^{-6}$ which is approximately one order of magnitude improvement compared to the previous ORCA demonstrations \cite{Kaczmarek2018,Finkelstein2018,Gao2019}, and to our knowledge the lowest reported for atom-based quantum memories. 

We can infer the quantum performance this memory could achieve. The second-order coherence of a retrieved single photon (which has ideally $\mu_\textrm{in} = 1$ and $g^{(2)}_\textrm{in} = 0$) is given by $g^{(2)}_\textrm{out} = 2/(\mu_\textrm{in}/\mu_\textrm{1}+1)$ \cite{Thomas2019}, under the assumption that the noise is thermal and is created independently of the memory process, so that the output can be treated as an incoherent admixture of signal and noise. Our ORCA memory would retrieve the photon with $g^{(2)}~=~9(1)\,\times\,10^{-6}$. Furthermore, the output fidelity for single-photon storage follows $F = (\mu_\textrm{in} + \mu_\textrm{1})/(\mu_\textrm{in} + 2\mu_\textrm{1})$ \cite{Gundogan2015}. The retrieved qubit fidelity of the ORCA memory would be at the $99.9996\,\%$ level.

\nit{Discussion} -- Our telecommunication-wavelength ORCA quantum memory has demonstrated exceptional SNR performance over a GHz bandwidth. This opens the possibility of interfacing with single-photon sources based on parametric-down conversion or four-wave mixing with modest spectral filtering to match our memory bandwidth, as well as to InGaAs quantum dot photon sources \cite{Kolatschek2021, Lettner2021} the bandwidths of which are well-matched to our device. Total memory efficiencies approaching unity should be achieved with optimal temporal-spectral mode shaping of the control pulse \cite{Kaczmarek2018,Gao2019,Guo2019}. For some applications, the limitation of our memory is the Doppler-induced storage time, as discussed previously. Magneto-optical trapping of the ensemble would eliminate atomic motion, albeit with the additional complexity of ultrahigh-vacuum chambers. An alternate approach to eliminate the dephasing is to perform velocity-selective optical pumping (VSP) to target atoms with a narrow spread of velocities \cite{Main2021} or even tailor atomic frequency comb structures for rephasing \cite{Main2021b}. These trapped and VSP approaches however sacrifice significant atom numbers and therefore total memory efficiency. An approach that utilises all atoms is to apply additional optical fields that compensate the dephasing by means of a controlled dynamic AC-Stark shift \cite{Finkelstein2021}, or using additional optical fields to map to an auxiliary state in order to facilitate a backward read out that would automatically undo the accumulated phase \cite{Moiseev2001}. With Doppler-dephasing eliminated, the storage time would be limited to the decay time of the $4$D$_{5/2}$ state, of around $90\,$ns. Overcoming this would require additional fields to map to longer-lived states, e.g. a higher-lying Rydberg state, or to ground state.

\nit{Conclusion} -- We present a GHz-bandwidth quantum memory for telecom C-band light. We demonstrate a total efficiency of $20.90(1)\,\%$ and a lifetime of $1.10(2)\,$ns. The noise performance is unprecedented with a measured SNR of  $1.9(1)\,\times\,10^{4}$ for $\mu_\textrm{in} = 0.084$, allowing for input states with an average photon number as low as $4.5(6)\times10^{-6}$ to yield SNR = 1. We discuss avenues for further improvements on both storage time and efficiency, which would fulfil a critical requirement for future quantum technologies. 

\nit{Acknowledgements} -- This work was supported by: Engineering and Physical Sciences Research Council via the Quantum Computing and Simulation Hub (T001062); Horizon 2020 research and innovation program under Grant Agreement No.899814 (Qurope). PML acknowledges a UKRI Future Leaders Fellowship Grant Reference MR/V023845/1.

\end{document}